\documentclass[12pt,preprint]{aastex}

\shorttitle{AIC \& MSPs}
\shortauthors{Ablimit }

\begin{document}


\title{ Accretion-induced Collapse from Magnetic White Dwarf Binaries and Formation of Binary Millisecond Pulsars: Redbacks and Black Widows}
\author{Iminhaji Ablimit\altaffilmark{1,\star}  }
\altaffiltext{1}{Department of Astronomy, Kyoto University, Kitashirakawa-Oiwake-cho, Sakyo-ku, Kyoto 606-8502, Japan;  iminhaji@kusastro.kyoto-u.ac.jp}

\altaffiltext{}{$^\star$JSPS International Research Fellow}


\begin{abstract}

Redbacks (RBs) and black widows (BWs) are two peculiar classes of eclipsing millisecond pulsars (MSPs). The accretion-induced collapse (AIC) of an oxygen/neon/magnesium composition white dwarf to a neutron star has been suggested as one possible formation pathway for those two classes of MSPs. However, it is difficult to produce all known MSPs
with the traditional AIC scenario. In this study by using the MESA stellar evolution code, we investigate the detailed pre-AIC evolution of magnetized white dwarf binaries with the magnetic confinement model where the high magnetic field strength of the white dwarf can confine the accreted matter in the polar caps. We find that the initial donor mass and orbital periods in our model can be lower than that of previous traditional AIC models. We also present post-AIC evolution models to form RBs and BWs with and without the spin down luminosity evaporation of MSPs. Under the magnetic confinement model and evaporative winds (with corresponding angular momentum loss from the surface of the donor star), the companion masses and orbital periods of all known RBs can be covered 
and a number of binaries can evolve to become BWs.

\end{abstract}


\section{Introduction}

Millisecond pulsars (MSPs) are radio pulsars with pulsation periods less than 30 ms, and most of them are found in binary systems. Two evolution pathways have been proposed for the formation of the MSPs: the recycling scenario of low-mass X-ray binaries (LMXBs) and accretion-induced collapse (AIC) of massive white dwarfs (WDs; Bhattacharya \& van den Heuvel 1991). In the first way, the donor star provides the mass and angular momentum via Roche-lobe overflow (RLOF) to the neutron star (NS), and let the NS spin up to the order of millisecond with the decayed magnetic field (Alpar et al. 1982). In the latter one, an ONeMg WD accretes mass from its donor star and grows in mass to the Chandrasekhar limit mass or coalesces with its WD companion (the merger way is not for the binary MSP), and the following process is the electron capturing which lets the WD collapse to produce a fast spinning NS (Nomoto \& Kondo 1991). The AIC scenario has important contributions to the formation of MSPs (Hurley et al. 2010; Zhu et al. 2015; Liu et al. 2018).

Redbacks (RBs) and black widows (BWs) are the two different MSPs which have orbital periods lower than one day, and the companion stars of the RBs have masses higher than 0.1$M_\sun$, while the BWs have the lower mass companions  ($< 0.1 M_\sun$) (Roberts 2013). Previous studies claimed some important key issues, such as the irradiation-induced cyclic mass transfer and thermal-viscous instability of the accretion disk, in the formation of RBs and BWs via the evolution of converging LMXBs (e.g., Benvenuto et al. 2014; Jia \& Li 2015). In the pre-AIC evolution, the soft X-ray radiation from an accreting WD may trigger a strong wind from its donor (the self-excited wind model, King \& van Teeseling 1998), and Ablimit \& Li (2015) investigated the model to produce the RBs. The evaporation plays significant role in the evolution of post-AIC and LMXBs (van den Heuvel \& van Paradijs 1988; Roberts 2013), in which the irradiated high-energy particles by radio pulsars evaporate their donor stars and cause the stellar wind when there are no mass transfer from the donors. Liu \& Li (2017) studied the formation of RBs and BWs through the AIC channel by considering the possible effects of the evaporation during the post-AIC evolution, and they showed the effect of the different evaporation efficiency and angular momentum loss (AML) mechanisms on the final outcomes (see also Jia \& Li 2016). However, those works still have difficulties to form known RBs and BWs. Besides, there is the missing or ignored physical parameter (i.e., the magnetic field) which still needs to be investigated further in the pre-AIC evolution.

In the WD binary (pre-AIC) evolution, the magnetic field may have a key role. Around 25\% of all known cataclysmic variables (CVs) are magnetic CVs (Ferrario et al. 2015). They are divided into nonmagnetic CVs ($<$ 1MG), intermediate polars ($\sim$1$-$10 MG) and polars ($>$10 MG) depending on the magnetic field strength of the WD, and the observed WD magnetic field strength so far is up to 230 MG (Schmidt et al. 1999). The fraction of magnetic CV can indeed be even higher (?33\%, and 28\% of them are polars) according to the recent result from GAIA DR2 (Pala et al. 2019). The magnetic WDs in symbiotic binaries and super-soft X-ray sources (SSS) have been discovered (Kahabka 1995; Sokoloski \& Beldstin 1999; Osborne et al. 2001). The mean mass of high-field magnetic WDs is $\sim 0.8 M_\sun$ which is significantly higher than the mean mass of nonmagnetic WDs, $\sim 0.6 M_\sun$ (Kepler et al. 2013). These observational results suggest that the magnetic field of the WD should be considered in the WD binary evolution.

In this work, we calculate detailed numerical calculations of the pre-AIC evolution including the magnetic confinement based on Ablimit \& Maeda (2019). We also present post-AIC binary evolution with and without the evaporation during RLOF to form RBs and BWs. In Section 2, we describe the calculation models of pre- and post-AIC evolutions. The calculation results are discussed in Section 3. The paper is closed in Section 4 with the conclusions.

\section{The calculation model}

We calculate the ONeMg WD binary evolution including main sequence (MS) stars with initial masses ranging from 0.8 to 8 $M_\sun$ by using the version 10000 of Modules for Experiments in Stellar Astrophysics (MESA; Paxton et al. 2015), and considering the magnetic confinement model. H abundance $X = 0.70$, He abundance $Y = 0.28$ and metallicity $Z = 0.02$ are taken for the donor star. The ONeMg WDs with the initial mass of 1.2 $M_\sun$ in those binaries are assumed to be highly magnetized with the initial magnetic field strengths of $1.1\times10^8$ G. After the ONeMg WDs reached the Chandrasekhar limit mass ($M_{\rm Ch} =$1.38$\rm M_\sun$) and collapsed to form NSs, we also calculate post-AIC evolutions with and without the evaporation.

The formula in Eggleton (1983) is used  for the effective Roche-lobe (RL) radius of the donor star, and the Ritter scheme (Ritter 1988) is adopted to calculate the mass transfer via RL overflow in the code.  The angular momentum loss caused by gravitational wave radiation (GR, Landau \& Lifshitz 1975)  and magnetic braking (MB with $\gamma =3.0$, Verbunt \& Zwaan 1981; Rappaport et al. 1983) is also included.The revised part in the binary evolutions are introduced as below, and other things in the evolutions are same as typical ones introduced in the instrumental papers of MESA (e.g., Paxton et al. 2015).

\subsection{The pre-AIC evolution with the magnetic confinement}
\label{sec:model}

The accretion disk formation is prevented by the high strength magnetic field of the WD (Cropper 1990).  When the RLOF mass transfer occurs, the stream of matter from the donor is captured by the magnetic field of the WD, then it follows the magnetic lines and falls down onto the magnetic poles of the WD as an accretion column (Figure 1).  Livio (1983) showed that the magnetic confinement happens with the high-field magnetic WD binary like polar system, and the critical value of magnetic field strength for the magnetic confinement depends on the mass of WD. To have the magnetic confinement, the magnetic field strength B of the WD satisfies the physical condition (Spitzer 1962; Livio 1983) below,

\begin{equation}
\rm{B} \geq 9.3\times10^{7} (\frac{\rm{R_{WD}}}{5\times10^8\,\rm cm})(\frac{\rm{P_b}}{5\times10^{19}\,\rm{dyne\,cm^{-2}}})^{7/10}(\frac{\rm{M_{WD}}}{\rm M_\sun})^{-1/2}(\frac{\dot{M}}{10^{-10}\,\rm{M_\sun\,yr^{-1}}})^{-1/2}\, Gauss,
\end{equation}
where $\dot{M}$ is the RLOF mass transfer. $\rm{P_b}$ is the pressure at the base of accreted matter. 
For the mass ($\rm{M_{WD}}$) -- radius ($\rm{R_{WD}}$) relation of the WD, the following equation is used (Nauenberg 1972),

\begin{equation}
\rm{R_{WD}} = 7.8\times10^8[(\frac{\rm{M_{WD}}}{\rm{M_{Ch}}})^{-2/3} - (\frac{\rm{M_{WD}}}{\rm{M_{Ch}}})^{2/3}]^{1/2} .
\end{equation}
 
With the condition (Eq. 1) shown above, the accreted matter can be confined to the polar column. Accreted matter can not diffuse perpendicularly to the magnetic field lines and spread over the WD's surface, and the magnetic field confinement can prevent the nova outburst (Livio 1983).
The relation of the following isotropic pole-mass transfer rate ($\dot{M}_{\rm{p}}$) with the the RLOF mass transfer rate ($\dot{M}$) is,

\begin{equation}
\dot{M}_{\rm{p}} =  \frac{\rm S}{\Delta \rm{S}} {\dot{M}},
\end{equation}
where $\rm S$ is the surface area of the WD, and $\Delta \rm{S}$ is that of the dipole regions. $\Delta \rm{S}$ can be calculated by the equations of dipole geometry and Alfv$\acute{\rm e}$n radius of the magnetic WD.


 The mass retention efficiency is one of the important things which causes different results in the accreting WD. The prescription of the Prialnik's group for the efficiency of hydrogen burning (see Hillman et al. 2016), and methods of Kato \& Hachisu (2004) is used for the mass accumulation efficiency of helium, then the mass growth rate of the WD is,

\begin{equation}
\dot{M}_{\rm{WD}} =  \eta_{\rm H} \eta_{\rm He} {\dot{M}},
\end{equation}
where $\eta_{\rm H}$ and $\eta_{\rm He}$ are efficiencies of hydrogen and helium burning. For the polar-like binaries, the accreted matter through RLOF can be confined in the polar caps due to the high magnetic field of the WD. Thus, the efficiencies become as a
function of $\dot{M}_{\rm{p}}$, and it should be compared to the critical mass transfer rate, rather than the usual RLOF mass transfer rate (Ablimit \& Maeda 2019).

\subsection{The post-AIC evolution}

It is assumed that the WD collapses to be an NS with a gravitational mass of 1.25 $M_\sun$ when $M_{\rm{WD}} = M_{\rm{Ch}}$, and a mass of 0.13 $M_\sun$ converts into the binding energy. That amount of mass loss would make the orbit wider and cause a temporary detachment of the RL. The relation between the orbital separations just before ($ a_0$) and after ($ a$) the collapse is (Verbunt et al. 1990),

\begin{equation}
\frac{a}{a_0} =  \frac{M_{\rm WD} + M_2}{M_{\rm NS} + M_2},
\end{equation}
where $M_{\rm{NS}}$ and $M_2$ are the NS and donor star masses. According to Hurley et al.(2010) and Tauris et al.(2013), we assume that the newborn NS would not give a significant kick, and the eccentric orbit by the mass loss during the AIC could be efficiently circularized by the tidal effect.
 
After a magnetic NS binary with the MS star formed, the RL over flow mass transfer occurs again due to the expansion of the donor or the orbital shrinkage by the MB. An accretion disk will be formed when the transferred matter surrounds the NS. The accretion disk can be stable when the mass transfer rate is higher than the critical value given as (Dubus et al. 1999),

\begin{equation}
\dot{M}_{\rm c} =  3.2\times10^{-9}(\frac{M_{\rm NS}}{1.4\rm M_\sun})( \frac{M_2}{1.0\rm M_\sun})
(\frac{P_{\rm orb}}{1.0\,\rm d})\,\rm M_\sun \rm{yr}^{-1},
\end{equation}
where $P_{\rm{orb}}$ is the orbital period. The mass transfer process is a nonconservative and Eddington limited, the NS accretion is as following, 

\begin{equation}
\dot{M}_{\rm NS} =  \rm{min}(-\beta \dot{M}, \dot{M}_{\rm Edd}),
\end{equation}
where $\dot{M}$ is the RLOF mass transfer from the donor to NS, and
\begin{equation}
\dot{M}_{\rm Edd} =  3.6\times10^{-8}(\frac{M_{\rm NS}}{1.4\rm M_\sun})( \frac{0.1}{GM_{\rm NS}/R_{\rm NS}c^2})(\frac{1.7}{1.0+X})\,\rm M_\sun \rm{yr}^{-1},
\end{equation}
X is the hydrogen abundance of the accreting matter and the NS radius $R_{\rm NS}$ is $10^6$ cm. We take $\beta = 0.5$, and the excess matter is assumed to be expelled by the NS in the isotropic wind carrying the specific angular momentum of the NS. The NS can spin up to periods of milliseconds by accreting sufficient mass ($\sim 0.1 M_\sun$).  There is a evaporation wind from the donor star when it is irradiated by MPSs. The evaporated mass loss rate can be (van den Heuvel \& van Paradijs 1988),

\begin{equation}
\dot{M}_{2,\rm ev} =  -\frac{f}{2{\upsilon}^2_{2,\rm esc}}L_{\rm p}( \frac{R_2}{a})^2,
\end{equation}
where $f$ is the efficiency of the MSP's irradiation, ${\upsilon}^2_{2,\rm esc}$ is the escape velocity of the donor star, $R_2$ is the radius of the donor star, and $L_{\rm p} = 4{\pi}^2 I\dot{P}/P^3$ is the spin down luminosity of the MSP. For the spin down luminosity caused by the magnetic dipole radiation, we take typical values such as the moment of inertia $I=10^{45} \,\rm{g\,cm^{2}}$, the spin period $P = 3 \,\rm{ms}$ and its derivative $\dot{P} = 10^{-20}\, \rm{s\,s^{-1}}$. The accretion disk will be thermally and viscously unstable when 
$\dot{M}<\dot{M}_{\rm c}$. The NS does not accrete any matter from the donor during the evaporation process. 

For the angular momentum loss (AML) during the post-AIC, it can be expressed as,
\begin{equation}
\dot{J} =  \dot{J}_{\rm GR} + \dot{J}_{\rm MB} + \dot{J}_{\rm ML} + \dot{J}_{\rm ML, ev},
\end{equation}
where is $\dot{J}_{\rm GR}$, $\dot{J}_{\rm MB}$ and $\dot{J}_{\rm ML}$ are the AML due to the GR, MB and mass loss from NS, respectively. $\dot{J}_{\rm ML, ev}$ is the AML by the evaporation wind. The evaporation wind takes specific angular momentum from the donor surface, and this causes the orbital shrinkage (see model A in Jia \& Li 2015). If no evaporation ($f=0.0$), then $\dot{J}_{\rm ML, ev}=0.0$.

\section{Results and discussion}

We found the initial donor masses of highly magnetized WD binaries with $M_{\rm WD, i} = 1.2 M_\sun$ for the pre-AIC are ranged from 0.8 to 3.2 $M_\sun$ with the initial orbital periods of 0.3 - 25 days while they are 2.2 - 3.4 $M_\sun$  and 0.4 - 4.5 days for the non-magnetic case (see Ablimit \& Maeda 2019). Thus, the lower limit of the initial donor mass ($M_{\rm 2, i} =\sim 0.8 M_\sun$) for the AIC under the magnetic confinement model is much lower than that of previous works (i.e. it is $\sim 2.0 M_\sun$ in Tauris et al. (2013); Liu \& Li (2017)). Thus, when the WD reaches the Chandrasekhar mass, the donor mass and orbital period in this work can be very low comparing to the results in the non-magnetic case (see also Liu \& Li (2017); Ablimit \& Maeda 2019). It can be possible to produce NS binaries with very low mass companions and short orbital periods via the pre-AIC pathway under the magnetic confinement model.

 In this section, we discuss detailed pre- and post-AIC evolutions of three examples. Then, the final mass and orbital period distributions of selected samples from our calculations are discussed and compared with that of observations.


 The detailed evolutions of three typical binaries with the magnetic confinement are demonstrated in Figure 2. The initial donor masses and orbital periods of the three binaries are 1.0 $M_\sun$ with 0.35 day (sample 1, S1), 1.6 $M_\sun$ with 4.0 days (sample 2, S2) and 2.2 $M_\sun$ with 3.4 days (sample 3, S3), respectively. It can be seen that the evolutions of S1 and S2 are like a CV evolution.  There will be a short time drop in the mass transfer in the CV evolution If the binary evolves into a period gap (see Paxton et al. 2015).  The RLOF mass transfer from the donors in S1 and S2 are lower than the mass transfer rate for the stable burning (Nomoto et al. 2007), so the nova burst would follow with those low mass donors. However, accreted hydrogen can burn stably in the polar caps and no nova burst happen when they become the pole-mass transfer (red lines in the upper panels of Figure 2) under the help of the magnetic confinement, and the WDs can grow in mass. The donor star in the S3 is massive enough to drive the WD to $M_{\rm Ch}$ with the non-magnetic case, and the magnetic confinement model still works for it (see the right panels of Figure 2). The timescale of the pre-AIC evolution of S1 is short, and the donor in S1 fills its RL easily due to the small initial orbit and AML by the MB. In the S2 and S3, it takes longer time to fill the RL because of the longer initial periods. The orbital periods of S2 and S3 are increased in the later phase of the evolutions, because the matter can not be accreted fully under the higher mass transfer rates (higher than comparing to that of S1), and the partial mass loss causes the orbit to expand with time (Figure 2).


The pre-AIC evolution under the magnetic confinement model can leave the low mass donor stars with various orbital periods for the post-AIC binaries. The donor masses and orbital periods of NS binaries born after the AIC from those three magnetic WD binaries are $\sim$0.8, 1.0 and $1.0 M_\sun$ with $\sim$0.3, 3.0 and 2.7 days (S1, S2 and S3), respectively. The post-AIC evolutions of those three samples without the evaporation are demonstrated in Figure 3. If there is no evaporation, the post-AIC evolutions are same as standard LMXB evolution (Tauris \& Savonije 1999). 
The orbital evolutions of three samples are dominated by the MB before the mass exchange starts. After the RLOF mass transfer begins, the NSs accretes enough mass to become MSPs.
In the early time of the RLOF in the S1, the mass transfer rate higher than the disk-c (red line) which is the critical value for disk instability, and NSs can have stable accretion disk. In the late time evolution, the mass transfer processes are unstable. S2 has unstable mass accretion, because the mass transfer rate is always lower than disk-c. Thus, the most of the transferred mass escape from the S2, and only some of them are accreted by the NS (still enough let the NS become an MSP). The AML dominated by the mass loss in S1 and S2 continue for the longer time, and the mass losses make orbits wider (Figure 3).  Comparing to NS binaries S1 and S3, the system of S2 loses more mass in a shorter time, and the AML due to that mass loss widen the orbit in a larger scale. In the later phase of S3, the mass loss causes the orbital period increment, then the AML dominated again by the MB shrinks the orbit (Figure 3).

Jia \& Li (2015) investigated the effect of the different evaporation efficiency parameter ($f$) on the evolution of LMXBs. The donor mass becomes smaller and mass transfer rate decreases due to the higher value of $f$, and the AML taken from the donor surface by the evaporation wind can be more sufficient (see also Jia \& Li 2016). This model is applied for the post-AIC evolutions by Liu \& Li (2017), and their results show that the more efficient evaporation and corresponding AML from the donor surface are needed to cover more observed RBs. Base on results of those works, we take $f=0.35$, and assume that the specific angular momentum is carried away from the donor surface by the evaporation wind. The calculation results are given in Figure 4. In the post-AIC evolution, the evaporation is coupling with the occurrence of RLOF. The evaporation oblates the donors, it decreases the RLOF mass transfer rate, and it dominates the late orbital evolutions of three samples. Comparing to the case of no evaporation (Figure 3), the final donor masses are lower, and final orbital periods are different when the evaporation is the part of evolutions (Figure 4). 


To produce RBs and BWs, we select some binaries which have the donors with masses from $\sim 0.8$ to 1.2 $M_\sun$, and different orbital periods. Figure 5 shows the comparison between those selected evolutionary tracks and the observed RBs and BWs\footnote{ Data are taken from A. Patruno's MSP catalog, https://apatruno.wordpress.com/about/millisecond-pulsar-catalogue/.}.  Although the magnetic confinement model is included in the pre-AIC evolution, it is still difficult to reproduce some RBs and BWs without the evaporation during the post-AIC evolution (see the left panel of Figure 5). The evolutionary tracks generally tend to evolve to two (wide or tight orbits) branches if orbital periods of binaries at the beginning of the second RLOF above or below the so-called bifurcation period (Pylyser \& Savonije 1988, 1989).
By taking the evaporation into account, the results cover all known RBs and a number of BWs (see the right panel of Figure 5 for the case of evaporation). 
In the previous similar works (e.g., Tauris et al. 2013), it is really hard that a low mass donor ($< \sim 2.0 M_\sun$) drive the WD to reach a mass of 1.38 $M_\sun$ and lead to AIC, unless some unusual mechanisms are invoked to accelerate the mass transfer (Ablimit \& Li 2015) or to make the mass accretion more efficient. Under the magnetic confinement model, the low mass donor can lead the WD to the AIC by the higher mass accretion efficiency, and the newborn NS can have the lower mass donor with the different orbital period. The decrease of mass and AML from the donor by the evaporation in the post-AIC evolution is also important for RBs and BWs. By considering the effect of the magnetic confinement and evaporation in the pre- and post-AIC, all known RBs are covered in this work, and a number of binaries can evolve to be BWs. Our results also show that the RBs can evolve to the BWs (Benvenuto et al. 2014, 2015). Forming all observed BWs is still hard with our current results. For a good representation of the population, the better way is the statistics such as the likelihood of forming RBs versus
BWs and associated birthrates once the evolution timescales and relative numbers for a real population
have been factored in. Performing the population synthesis also may show the formation pathways to all known BWs.

\section{Conclusions}

It has been claimed that it is hard to form all known RBs and BWs by the traditional model, and some special scenarios should be considered (e.g., Ablimit \& Li 2015).
In this paper, we calculate the pre-AIC evolution (the ONeMg WD binary evolution) under the magnetic confinement model (Ablimit \& Maeda 2019). The transferred mass from the MS star can be confined and burn stably under the high enough magnetic field strength of the WD, and it increases the local accretion rate of the WD. Because of the effective accretion rate under the magnetic confinement, the lower limit for the initial mass of the donor star which let the WD can reach the Chandrasekhar limit mass can be lower, and it means that the mass of the newborn NS's companion star after the AIC can be lower than that of previous works (e.g., Tauris et al. 2013).

We also present the different post-AIC evolution tracks in the evaporation and no evaporation cases. The newborn NS can spin up to be an MSP after accreting enough mass, then the evaporation takes place. The evaporation wind decreases the donor mass and the RLOF mass transfer rate, and it takes the angular momentum from the donor surface in the evolution. The lower mass donor (from the pre-AIC), and oblation plus AML by the evaporation (in the post-AIC) enable our results to overlap the companion mass - orbital period distributions of all known RBs and part of observed BWs. 
Liu \& Li (2017) assumed that the NS from the AIC model is born as an MSP due to the rapid rotation of the accreting WD in the pre-AIC evolution. However, it is hard to claim that all newborn NSs in the AIC model could be MSPs, and Yoon \& Langer (2005) discussed the conditions for the direct formation of MSPs by the AIC. If an MSP could be directly born from the AIC, the evaporation could occur right after the AIC, and considerably reduce the mass of the companion star. By combining the magnetic confinement model and the directly newborn MSP for the pre- and post-AIC evolutions, it may be easier to reproduce all observed RBs, and possible to form more known BWs. In the future, it is also interesting to calculate a real population (i.e. statistics) under different assumptions and models for the AIC channel to produce wider parameter space for short-period binary millisecond pulsars, especially for BWs.

\begin{acknowledgements}

I thank the anonymous referee for the careful review of this work. This work was funded by JSPS International Postdoctoral fellowship of Japan (P17022), JSPS KAKENHI grant no. 17F17022. 

\end{acknowledgements}

\begin{center}
REFERENCES 
\end{center}

Ablimit, I., \& Li, X.-D. 2015, ApJ, 800, 98

Ablimit, I., \& Maeda, K. 2019, ApJ, 871, 31

Alpar, M. A., Cheng, A. F., Ruderman, M. A., \& Shaham, J. 1982, Nature,
300, 728

Benvenuto, O. G., De Vito, M. A., \& Horvath, J. E. 2014, ApJL, 786, L7

Bhattacharya, D., \& van den Heuvel, E. P. J. 1991, PhR, 203, 1

Dubus, G., Lasota, J.-P., Hameury, J.-M., \& Charles, P. 1999, MNRAS,
303, 139

Eggleton, P. P. 1983, ApJ, 268, 368

Ferrario, L., de Martino, D., \& Gnsicke, B. T. 2015, SSRv, 191, 111F
Hillman, Y., Prialnik, D., Kovetz, A., \& Shara, M. M. 2016, ApJ, 819, 168

Hurley, J. R., Tout, C. A., Wichramasinghe, D. T., Ferrario, L., \& Kiel, P. D.
2010, MNRAS, 402, 1437


Jia, K., \& Li, X.-D. 2015, ApJ, 814, 74

Jia, K., \& Li, X.-D. 2016, ApJ, 830, 153

Kato, M., \& Hachisu, I. 2004, ApJL, 613, L129

Kahabka, P. 1995, ASP Conference Series, Vol. 85

Kepler, S. O., Pelisoli, I., Jordan, S., et al. 2013, MNRAS, 429, 2934

King, A. R., \& van Teeseling, A. 1998, A\&A, 338, 965

Landau, L. D., \& Lifshitz, E. M. 1975, The Classical Theory of Fields (4th ed.;
Oxford: Pergamon)

Liu, D., Wang, B., Chen, W., Zuo, Z. \& Han, Z. 2018, MNRAS, 477, 384

 Liu, Wei-Min, \& Li, X.-D. 2017, ApJ, 851, 58
 
Livio, M. 1983, A\&A, 121, L7 

Nauenberg, M. 1972, ApJ, 175, 417

Nomoto, K., \& Kondo, Y. 1991, ApJL, 367, L19

Osborne, et al. 2001, A\&A, 378, 800

Paxton, B., Marchant, P., Schwab, J., et al. 2015, ApJS, 220, 15

Pala, A. F., G$\ddot{a}$nsicke, B. T., Breedt, E., et al., 2019, arXiv:1907.13152

Pylyser, E., \& Savonije, G. J. 1988, A\&A, 191, 57

Pylyser, E. H. P., \& Savonije, G. J. 1989, A\&A, 208, 52

Rappaport, S., Verbunt, F., \& Joss, P. C. 1983, ApJ, 275, 713

Ritter, H. 1988, A\&A, 202, 93

Roberts, M. S. E. 2013, in IAU Symp. 291, Neutron Stars and Pulsars:
Challenges and Opportunities after 80 Years, ed. J. van Leeuwen
(Cambridge: Cambridge Univ. Press), 127

Schmidt, G. D., Hoard, D. W., Szkody, P., Melia, F., Honeycutt, R. K., \& Wagner, R. M. 1999, ApJ, 525, 407

Sokoloski, J. L., \& Beldstin, L. 1999, ApJ, 517, 919

Spitzer, L. 1962, Physics of Fully Ionized Gases, (New York: John Wiley and Sons)
Tauris, T. M., Sanyal, D., Yoon, S.-C., \& Langer, N. 2013, A\&A, 558, A39

Tauris, T. M., \& Savonije, G. J. 1999, A\&A, 350, 928

van den Heuvel, E. P. J., \& van Paradijs, J. 1988, Natur, 334, 227

Verbunt, F., \& Zwaan, C. 1981, A\&A, 100, L7

Verbunt, F., Wijers, R. A. M. J., \& Burn, H. M. G. 1990, A\&A, 234, 195

Yoon S.-C., Langer N., 2005, A\&A, 435, 967

Zhu, C., L$\ddot{u}$, G., \& Wang, Z. 2015, MNRAS, 454, 1725



\begin{figure}
\centering
\includegraphics[totalheight=2.5in,width=4.8in]{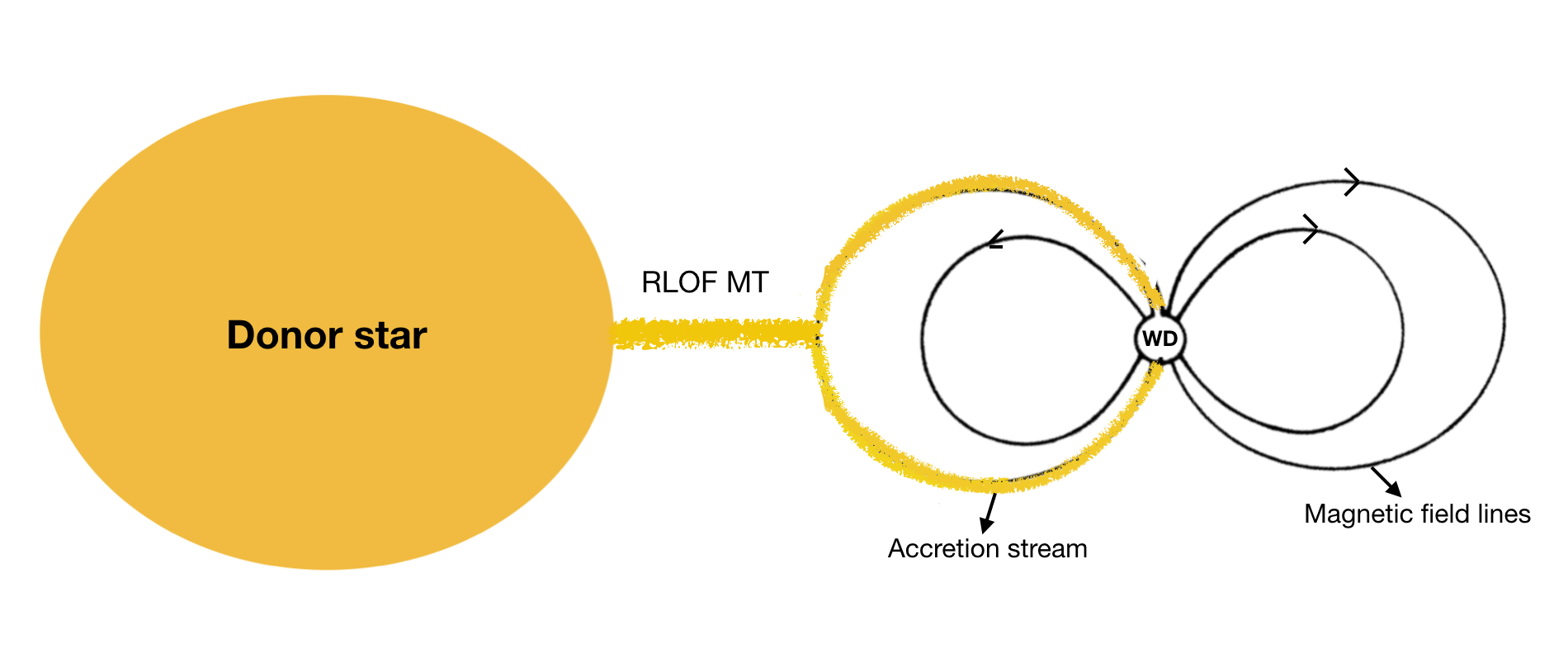}
\caption{Stream-like accretion along the magnetic lines of the WD in the highly magnetized WD binary.}
\label{fig:1}
\end{figure}

\clearpage



\begin{figure}
\centering
\includegraphics[totalheight=6.5in,width=7.5in]{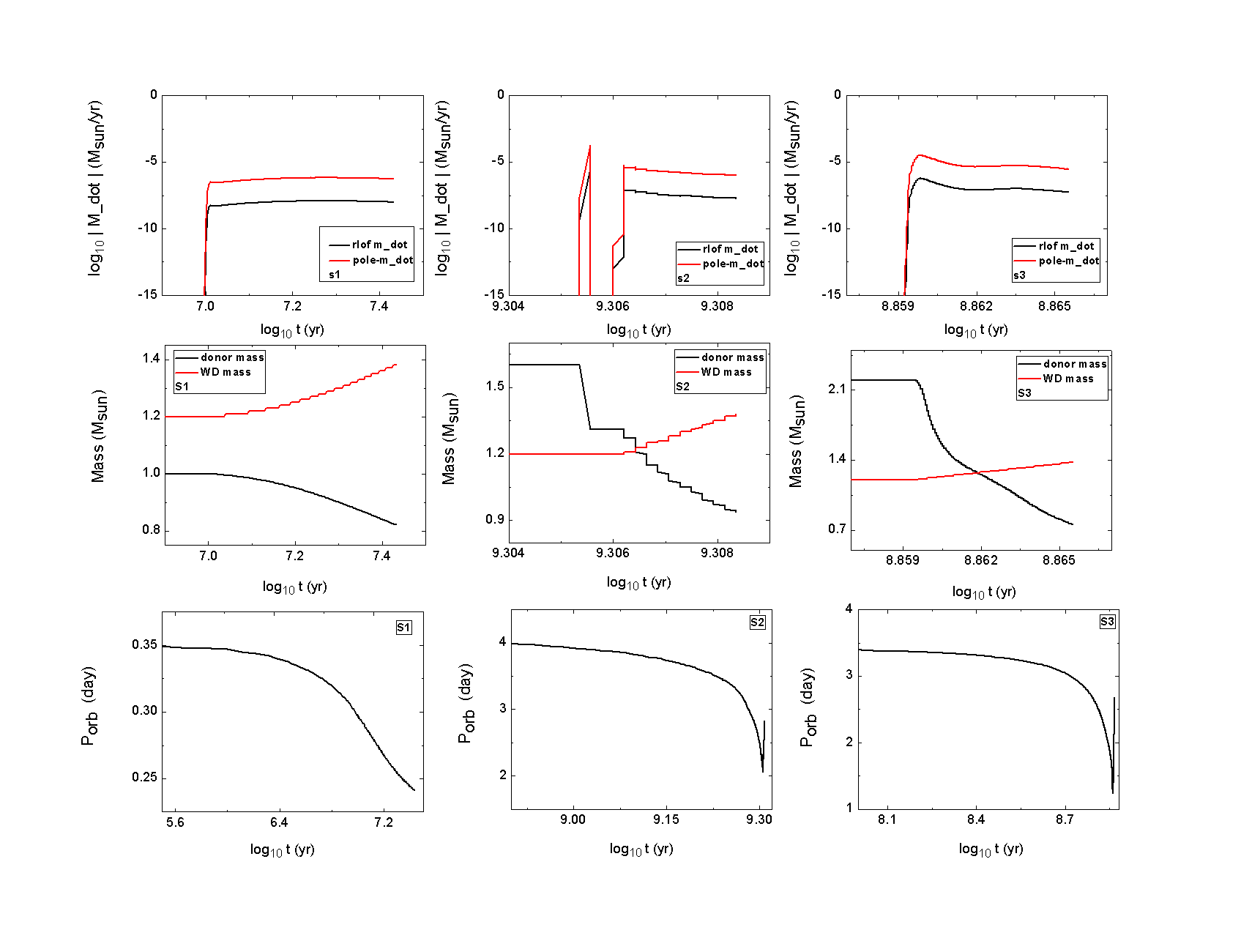}
\caption{The pre-AIC evolutions of three ONeMg WD binaries (S1, 2 and 3) under the magnetic confinement model. The initial masses and initial orbital periods of S1, S2 and S3 are 1.0, 1.6 and 2.2 $M_\sun$ with 0.35 day (left panels), 4.0 days (middle panels) and 3.4 days (right panels), respectively. The evolutions of the mass transfer rates, masses of WDs \& donor stars, and orbital periods with time are given.}
\label{fig:1}
\end{figure}

\clearpage


\begin{figure}
\centering
\includegraphics[totalheight=6.5in,width=7.5in]{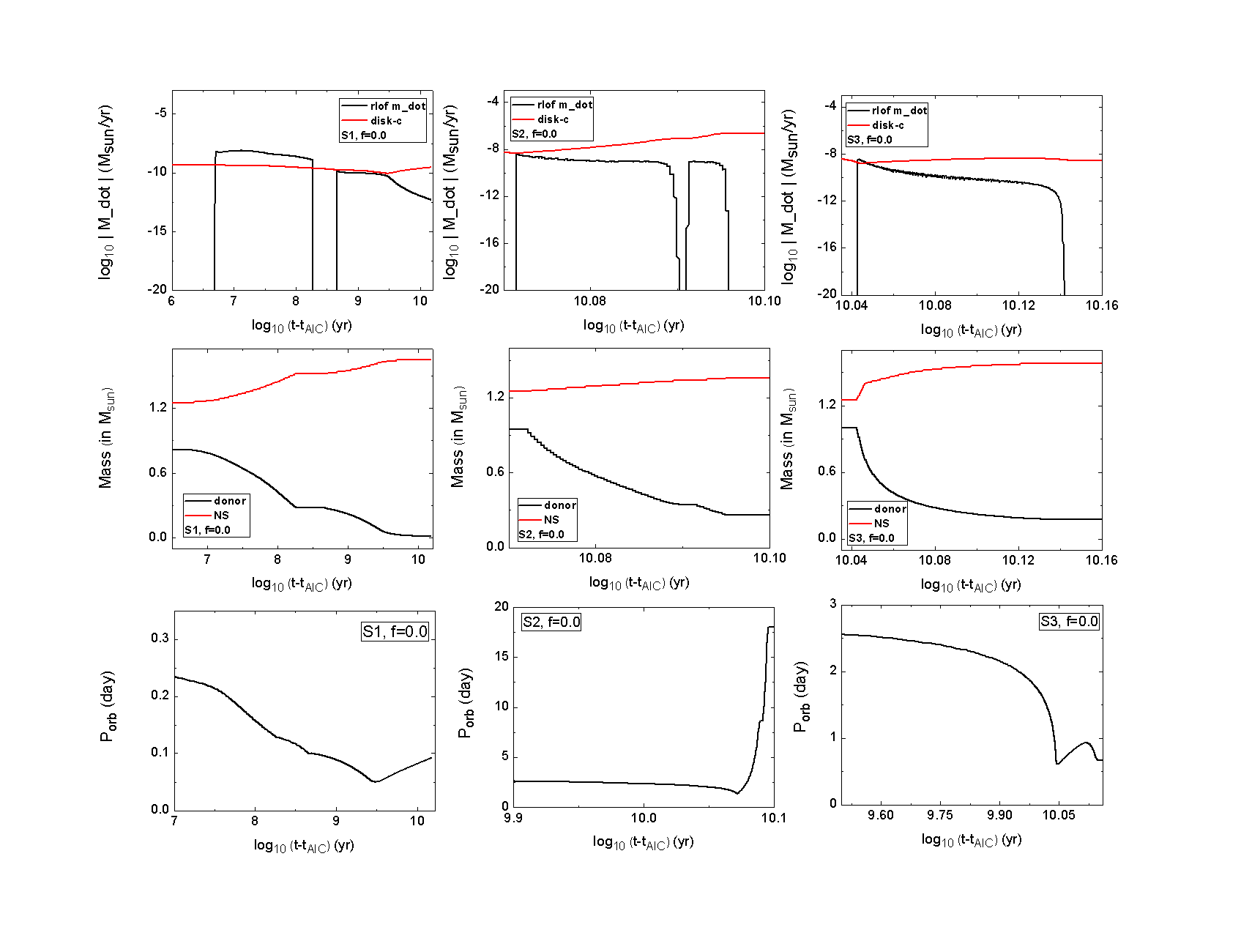}
\caption{Detailed post-AIC evolutions of three samples (S1,2 and 3) with $f=0.0$.   }
\label{fig:1}
\end{figure}

\clearpage

\begin{figure}
\centering
\includegraphics[totalheight=6.5in,width=7.5in]{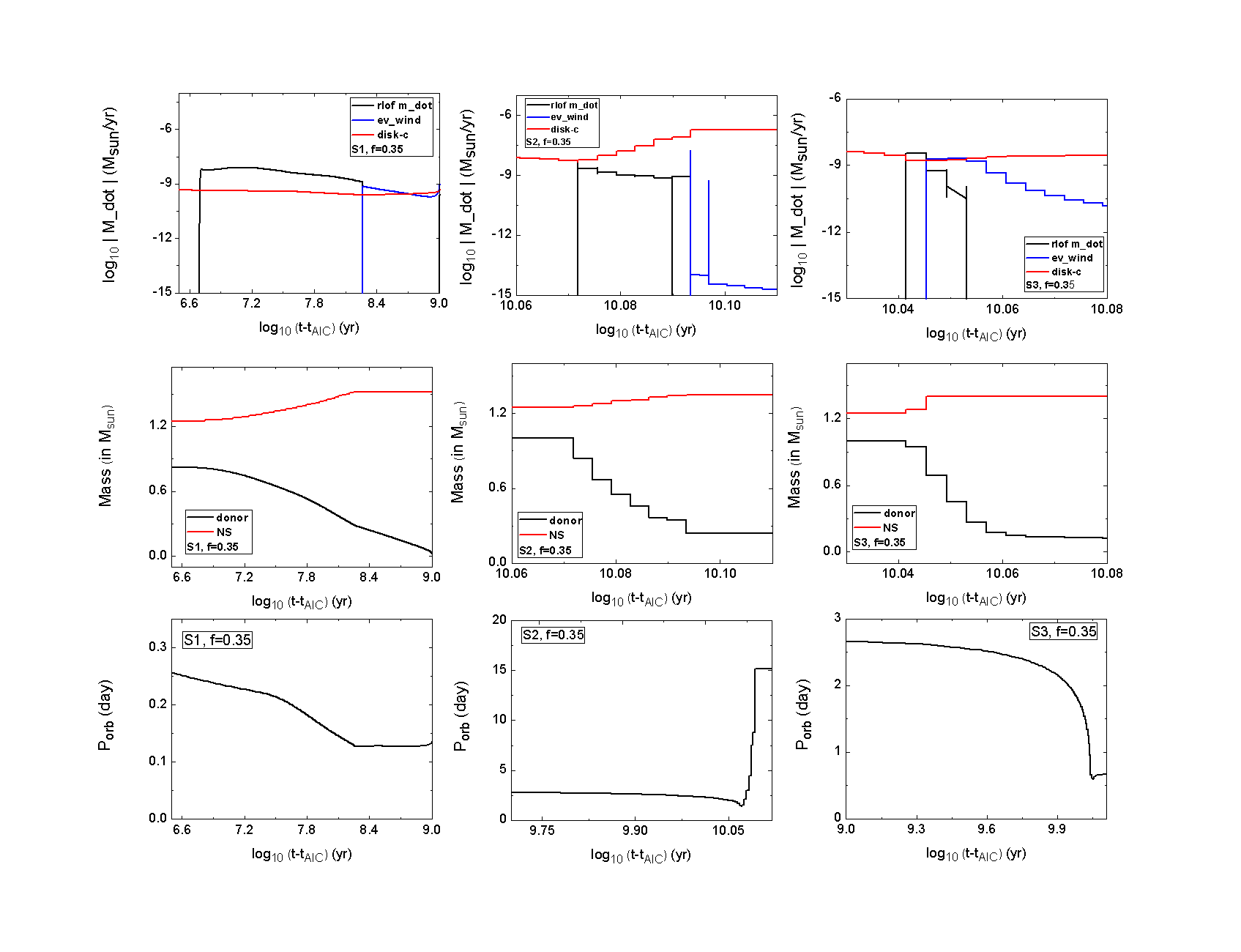}
\caption{Detailed post-AIC evolutions of three (S1,2 and 3) samples with $f=0.35$}
\label{fig:1}
\end{figure}

\clearpage

\begin{figure}
\centering

\includegraphics[totalheight=2.9in,width=3.2in]{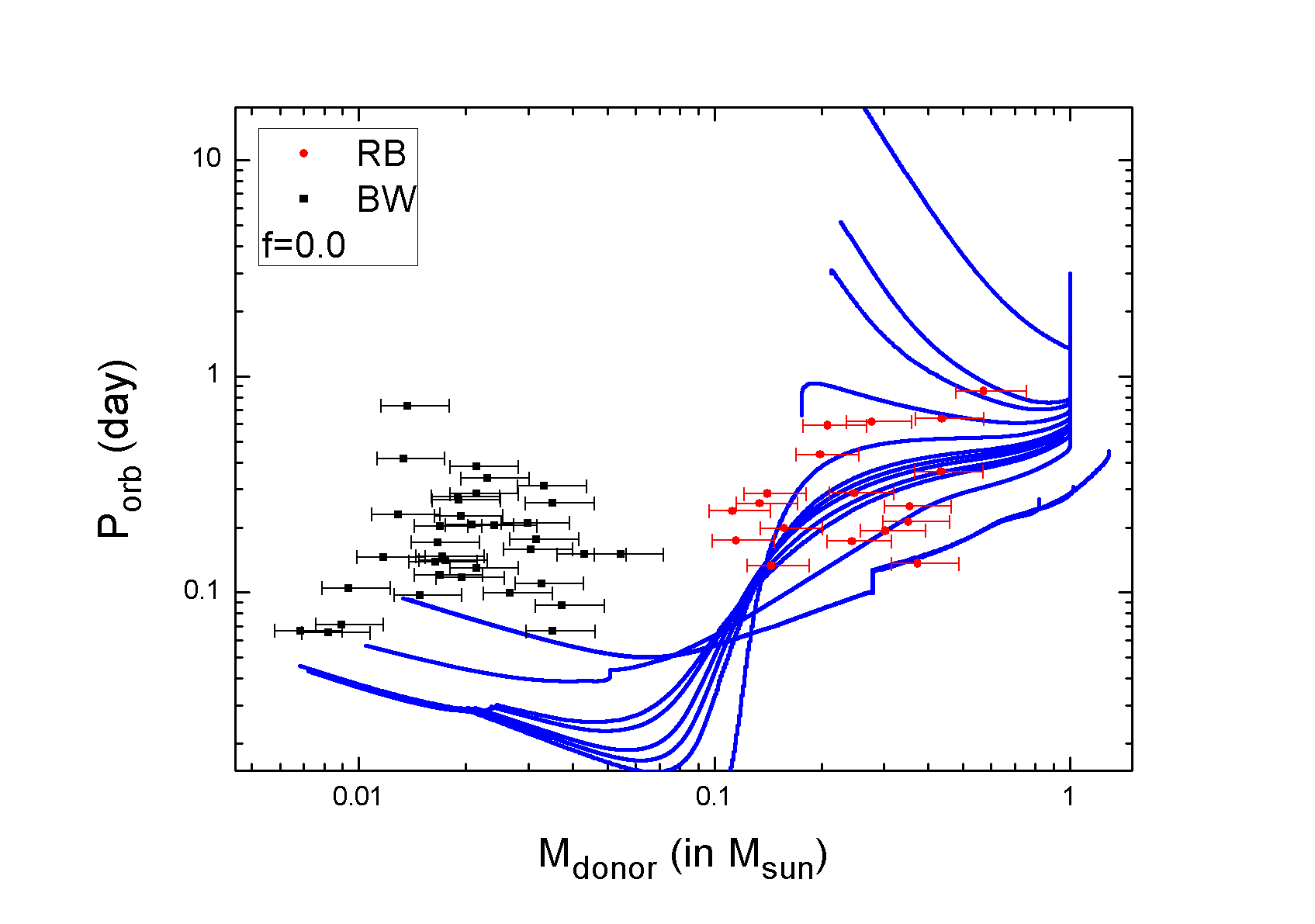}
\includegraphics[totalheight=2.9in,width=3.2in]{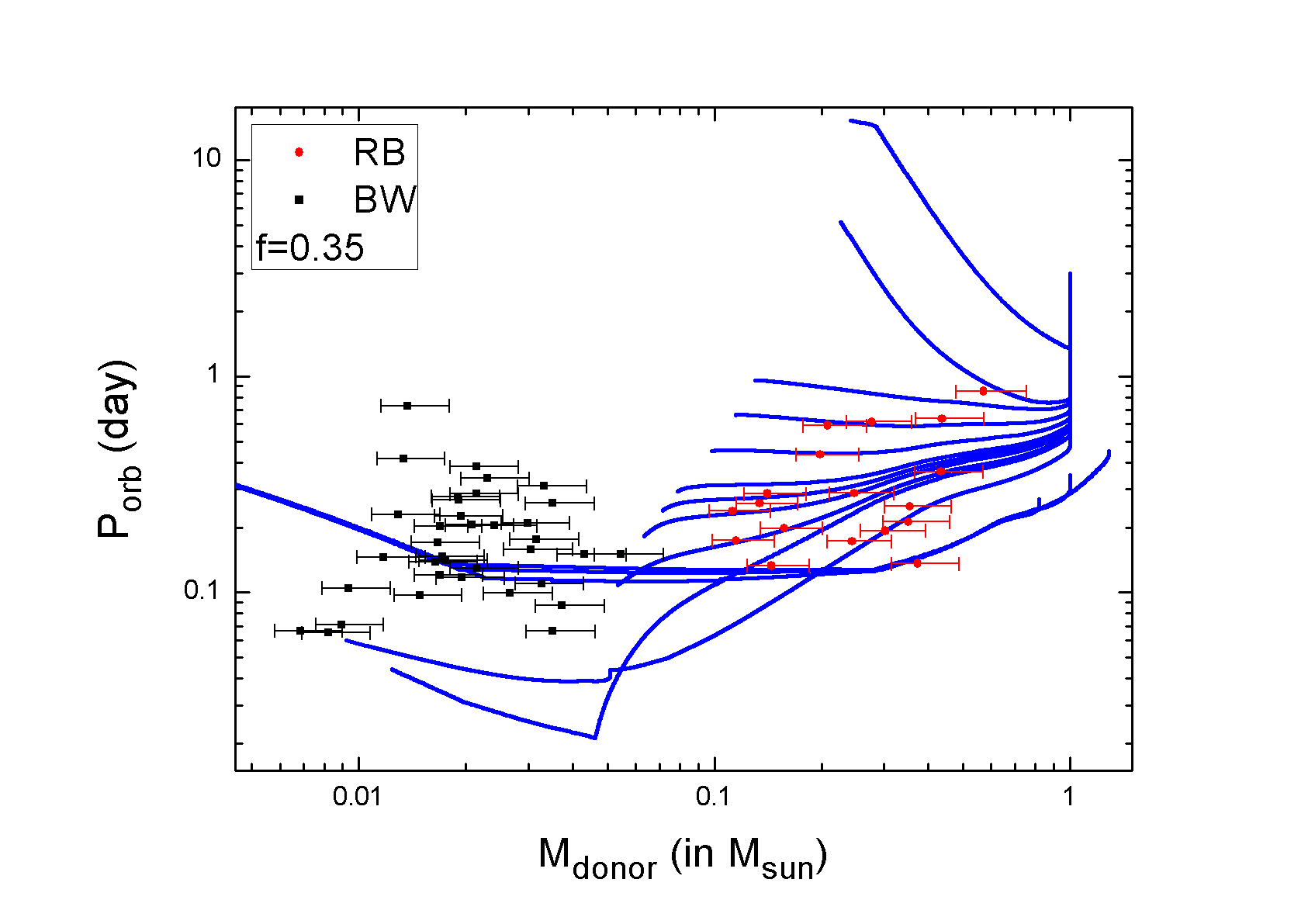}
\caption{Comparison of calculated evolutionary tracks of the post-AIC binaries with observed redbacks (red circles) and black widows (black squres) }
\label{fig:1}
\end{figure}

\clearpage

\end{document}